\documentclass[reprint,aps]{revtex4-1}

\usepackage{amsmath,latexsym,xcolor,mathtools}
\usepackage{graphicx,verbatim, subcaption, caption}
\usepackage{color}

\usepackage[title]{appendix}

\newcommand{\be}{\begin{equation}}
\newcommand{\ee}{\end{equation}}

\begin{document}
\title{Null hypersurfaces in Kerr-(A)dS spacetimes}
	\author{Abdulrahim Al Balushi}
	\email[]{a2albalu@uwaterloo.ca}
	\affiliation{Department of Physics and Astronomy, University of Waterloo, Waterloo, Ontario, N2L 3G1, Canada}
	\author{Robert B. Mann}
	\email[]{rbmann@uwaterloo.ca}
	\affiliation{Department of Physics and Astronomy, University of Waterloo, Waterloo, Ontario, N2L 3G1, Canada}
	 
\vskip -0.5cm 
\begin{abstract}
 For many purposes, a three-dimensional foliation of spacetime is more advantageous to understanding its light cone structure. We derive the equations describing such foliations for the Kerr geometry with non-zero cosmological constant, and show that they reduce to null hypersurfaces in vacuum (anti-)de Sitter spacetime in the limit of zero mass. Furthermore, we find that these null hypersurfaces are free of caustics everywhere for $r>0$. Our construction has applications in numerical studies of rotating black holes, and in defining Kruskal coordinates for rotating black holes with non-zero cosmological constant.
\end{abstract}

\pacs{Valid PACS appear here}
\maketitle


The discovery of the Kerr solution in 1963 \cite{Kerr1963} and its extension to the Kerr-Newman \cite{Newman1965} family of solutions of Einstein equations have allowed accurate analytical studies of astrophysical black holes. These solutions furthermore have opened the door for theoretical studies of rotating black hole thermodynamics \cite{Gunasekaran2012,Altamirano2014}, and quantum gravity within the context of the AdS/CFT correspondence \cite{Hawking:1998kw,Hawking2000} and the Kerr/CFT correspondence \cite{Bredberg:2011hp}. 

Null hypersurfaces of the Kerr geometry were first systematically studied in \cite{Pretorius1998}, where a three-dimensional null slicing of the spacetime was obtained and its properties studied. These hypersurfaces were found to possess no caustics, which make them ideal for studying initial-value problems and wave propagation in Kerr geometry.  

This type of analysis has never been extended to other rotating black holes, and it is the purpose
of this paper to address this deficit by investigating null hypersurfaces for the Kerr-Anti de Sitter
(Kerr-AdS) and Kerr-de Sitter (Kerr-dS) black holes. 
We find that a similar three-dimensional null foliation of Kerr-(A)dS spacetimes can be obtained and prove that it also develops no caustics. Furthermore, we comment on the difference in the behaviour of light cones between rotating AdS, flat, and dS geometries. 
Besides the desire for completeness, one application of these results motivating this study is that of  providing a better understanding of the geometry of null boundaries of ``Wheeler-de Witt" (WDW) patches in rotating AdS spacetime within the context of ``complexity equals action" conjecture \cite{Brown2016a,Brown2016} in AdS/CFT. 

 Our paper is organized as follows: in section I, we provide the most general solution for the null hypersurfaces $t\pm r_*=\text{const}$ in Kerr-(A)dS spacetimes in terms of elliptic integrals. In section II, the $m\rightarrow 0$ limit of the solution is taken and shown to correspond to light cones in (A)dS spacetimes. In section III, a three-dimensional foliation of the Kerr-(A)dS geometry is obtained and shown in section IV to possess no caustics for $r>0$. Finally, section V construct the Kruskal coordinates for the Kerr-(A)dS spacetime.  
\section{Preliminaries} 
The Kerr-(A)dS metric of the (3+1)-dimensional rotating black hole in Boyer-Lindquist coordinates is
\begin{widetext}
\begin{align}
ds^2&=-\frac{\Delta_r}{\Sigma^2}{\left(dt-\frac{a}{\Xi}\sin^2\theta\ d\phi\right)}^2+\frac{\Sigma^2}{\Delta_r}dr^2+\frac{\Sigma^2}{\Delta_{\theta}}d\theta^2+\frac{\Delta_{\theta}}{\Sigma^2}\sin^2\theta\ {\left(a\ dt-\frac{r^2+a^2}{\Xi}d\phi\right)}^2
\label{Kerr}
\end{align}
where
\begin{equation}
\Delta_r=\left(r^2+a^2\right)\left(1+\frac{r^2}{\epsilon L^2}\right)-2mr,\quad \Xi=1-\frac{a^2}{\epsilon L^2},\quad \Delta_{\theta}=1-\frac{a^2\cos^2\theta}{\epsilon L^2},\quad\Sigma^2=r^2+a^2\cos^2\theta
\end{equation}
where $\epsilon=+1$ for AdS and $\epsilon=-1$ for dS spacetimes, with $\Delta_r (r_+) =0$ defining
the outer horizon $r_+$ of the black hole.  The inner and cosmological horizons are respectively
defined from $\Delta_r (r_-) =0$ and $\Delta_r (r_c) =0$, the latter being present only for $\epsilon = -1$.
The rotation parameter is bounded by $a<L$. The relevant thermodynamic quantities are
\begin{equation}
M=\frac{m}{G_N\Xi^2},\ J=\frac{ma}{G_N\Xi^2},\ \Omega_{\pm}=\frac{a\left(1+\frac{r_{\pm}^2}{\epsilon L^2}\right)}{r_{\pm}^2+a^2},\ T=\frac{r_+}{2\pi}\left(1+\frac{r_+^2}{\epsilon L^2}\right)\frac{1}{r_+^2+a^2}-\frac{1}{4\pi r_+}\left(1-\frac{r_+^2}{\epsilon L^2}\right),\ S=\frac{\pi r_+^2}{4G_N}
\label{Kerrthermo}
\end{equation}
and are respectively its mass, angular momentum, horizon angular velocity, temperature, and entropy 
The metric is regular everywhere away from the symmetry axis for AdS, while it is regular and static only for $r<L$ in dS. Below, we will assume that $r>0$ for AdS and $r<L$ for dS, unless otherwise stated (i.e. in section 4).
\end{widetext} 
 
We want to find the null boundary surfaces $\Phi(x)=\text{const}$ in the spacetime \eqref{Kerr}. Define the ingoing/outgoing Eddington-Finkelstein coordinates by
\begin{equation}\label{nullgen}
v=t+ r_*,\qquad u=t-r_*
\end{equation}
The condition of the surfaces defined by $v=\text{const}$ being null translates to
\begin{align}
g^{\alpha\beta}\partial_{\alpha}v\partial_{\beta}v =
g^{tt}+g^{rr}{\left(\partial_rr_*\right)}^2+g^{\theta\theta}{\left(\partial_{\theta}r_*\right)}^2&=0
\label{r*}
\end{align}
Thus, the problem of finding the null hypersurfaces reduces to solving the PDE \eqref{r*} for $r_*(r,\theta)$. For the metric \eqref{Kerr}, 
\begin{align}
g^{tt}&=\frac{g_{\phi\phi}}{g_{tt}g_{\phi\phi}-g_{t\phi}^2}=-\frac{\Xi^2\left[-a^2\Delta_r\sin^2\theta+\Delta_{\theta}{(r^2+a^2)}^2\right]}{\Delta_r\Delta_{\theta}\Sigma^2}
\label{gtt}
\end{align}
yielding
\begin{align}
\Delta_r{\left(\partial_rr_*\right)}^2+\Delta_{\theta}{\left(\partial_{\theta}r_*\right)}^2 
&=\Xi^2\left[\frac{{(r^2+a^2)}^2}{\Delta_r}-\frac{a^2\sin^2\theta}{\Delta_{\theta}}\right]
\label{PDE}
\end{align}
for the PDE \eqref{r*}. 
In the limit $L\rightarrow \infty$  this reduces to the  asymptotically flat case \cite{Pretorius1998}
\begin{equation}
\Delta{\left(\partial_rr_*\right)}^2+{\left(\partial_{\theta}r_*\right)}^2=\frac{{(r^2+a^2)}^2}{\Delta}-a^2\sin^2\theta
\end{equation}
where now $\Delta\equiv\lim\limits_{L\rightarrow\infty}\Delta_r=r^2+a^2-2mr$. 

This separable form allows us to easily guess an ansatz for $r_*(r,\theta)$.  First, define 
\begin{align}
Q^2(r) &=\Xi^2\left[{(r^2+a^2)}^2-a^2\lambda\Delta_r\right] \nonumber \\
P^2(\theta) &=\Xi^2a^2\left[\lambda\Delta_{\theta}-\sin^2\theta\right]
\label{PQ}
\end{align}
where $\lambda$ is an arbitrary constant. Then, it is clear that choosing
\begin{equation}
\partial_rr_*=\frac{Q(r)}{\Delta_r},\qquad \partial_{\theta}r_*=\frac{P(\theta)}{\Delta_{\theta}}
\end{equation}
would satisfy \eqref{PDE}. Hence a solution to \eqref{PDE} is obtained by solving the exact integral
\begin{align}
dr_*&=\frac{Q}{\Delta_r}dr+\frac{P}{\Delta_{\theta}}d\theta
\label{r**}
\end{align}

To find a general solution $r_*(r,\theta)$ of \eqref{PDE} that is independent of $\lambda$ we follow the procedure in  \cite{Pretorius1998} and assume first that $\lambda$ is now a function of $r$ and $\theta$. In this case, $r_*=\rho(r,\theta,\lambda)$ where
\begin{equation}
d\rho=\frac{Q}{\Delta_r}dr+\frac{P}{\Delta_{\theta}}d\theta+\frac{a^2}{2}Fd\lambda
\label{rho}
\end{equation}
where $\partial_\lambda\rho(r,\theta,\lambda)=\frac{a^2}{2}F(r,\theta,\lambda)$, and
\begin{equation}
F(r,\theta,\lambda)=\int_r^\infty\frac{1}{Q(r',\lambda)}dr'+\int_0^\theta\frac{1}{P(\theta',\lambda)}d\theta'+g'(\lambda)
\label{F}
\end{equation}
The condition \eqref{r**} implies that 
\begin{equation}
F(r,\theta,\lambda)=0
\label{F0}
\end{equation}
which fixes the dependence of $\lambda$ on $(r,\theta)$ for any given choice of the function
 $g[\lambda(r,\theta)]$.  The explicit form of the general solution of \eqref{rho} is then
\begin{equation}
\rho(r,\theta,\lambda)=\int_0^r\frac{Q}{\Delta_r}dr+\int_0^\theta\frac{P}{\Delta_{\theta}}d\theta+\frac{a^2}{2}g(\lambda)
\label{rho0}
\end{equation}

 Once $g(\lambda)$ is chosen, the exact integrals in \eqref{rho0} and \eqref{F0} are performed assuming that $\lambda$ is a constant. Then, \eqref{F0} is used to solve for $\lambda(r,\theta)$, which in turn is substituted into the result obtained upon integrating \eqref{rho0}.  The net result is that $r_*(r,\theta)=\rho(r,\theta,\lambda(r,\theta))$ can be explicitly obtained.

\section{$m\rightarrow 0$ limit: Light cones in vacuum $\text{(A)dS}$ metric}

Here, we verify the expressions for the light cones above by taking the $m\rightarrow 0$ limit and showing that they reduce to light cones in vacuum (A)dS spacetime. We begin by  
simplifying   the metric \eqref{Kerr} using the
coordinate transformation \cite{Agnese:1999df}\begin{equation}
t\rightarrow \Xi t,\qquad \phi\rightarrow\Xi^{1/2}\phi-\frac{a\Xi}{\epsilon L^2}t
\end{equation} 
yielding
\begin{align}
ds^2&=-\frac{\Delta_\theta}{\Sigma^2}\left[\Delta_r\Delta_\theta-a^2\sin^2\theta{\big(1+\frac{r^2}{L^2}\big)}^2\right]dt^2+\frac{\Sigma^2}{\Delta_r}dr^2\nonumber\\
&\qquad +\frac{\Sigma^2}{\Delta_\theta}d\theta^2+\frac{\sin^2\theta}{\Sigma^2\Xi}\left[{(r^2+a^2)}^2\Delta_\theta-a^2\Delta_r\sin^2\theta\right]d\phi^2 \nonumber\\
&\qquad\qquad -\frac{4mra\Delta_\theta\sin^2\theta}{\Sigma^2\Xi^{1/2}}dtd\phi
\end{align}
Taking the $m\rightarrow 0$ limit now gives
\begin{align}
ds^2 &= -\Xi\Delta_\theta \left(1+\frac{r^2}{\epsilon L^2}\right)dt^2+\frac{\Sigma^2}{\left(r^2+a^2\right)\left(1+\frac{r^2}{L^2}\right)}dr^2\nonumber\\
&\qquad +\frac{\Sigma^2}{\Delta_\theta}d\theta^2+\left(r^2+a^2\right)\sin^2\theta d\phi^2
\label{KerrAdSLimit}
\end{align}
This metric is just the vacuum (A)dS metric 
\begin{equation}
ds^2=-\left(1+\frac{r^2}{\epsilon L^2}\right)dt^2+{\left(1+\frac{r^2}{\epsilon L^2}\right)}^{-1}dr^2+r^2d\theta^2+r^2\sin^2\theta d\phi^2
\label{vacAdS}
\end{equation}
where the coordinate transformation from \eqref{vacAdS} to \eqref{KerrAdSLimit} is given by \cite{Gibbons2017} 
\begin{align}
\label{coordtrans1}
t&\rightarrow \Xi t\\
r&\rightarrow \sqrt{\frac{r^2\left(1-\frac{a^2\cos^2\theta}{\epsilon L^2}\right)+a^2\sin^2\theta}{\Xi}}\\
\theta&\rightarrow \cos^{-1}\left(\frac{r\sqrt{\Xi} \cos\theta}{\sqrt{r^2\left(1-\frac{a^2\cos^2\theta}{\epsilon L^2}\right)+a^2\sin^2\theta}}\right)\\
\phi&\rightarrow \sqrt{\Xi}\phi
\label{coordtrans2}
\end{align}

In the limit $m\rightarrow 0$,
\begin{equation}
Q\rightarrow Q_0=\Xi\sqrt{\left(r^2+a^2\right)\left[r^2+a^2-a^2\lambda\big(1+\frac{r^2}{\epsilon L^2}\big)\right]}
\end{equation}
and $P$ remains the same. To simplify the integrals, we make the substitution 
\begin{equation}
r=\frac{a\sqrt{1-\lambda}\sin\chi}{\sqrt{\lambda\Delta_\chi-\sin^2\chi}}
\label{r}
\end{equation}
where $\Delta_\chi\equiv\lim\limits_{m\rightarrow 0}\Delta_r$, from which it follows that
\begin{equation}
\frac{dr}{Q_0}=\frac{d\chi}{a\Xi\sqrt{\lambda-\sin^2\chi-\frac{a^2\lambda\cos^2\chi}{\epsilon L^2}}}
\end{equation}
The function $F$ in \eqref{F} becomes
\begin{widetext}
\begin{align}
F(r,\theta,\lambda)&=\int_{\chi(r,\lambda)}^{\theta^*}\frac{d\chi'}{a\Xi\sqrt{\lambda-\sin^2\chi'-\frac{a^2\lambda\cos^2\chi'}{\epsilon L^2}}}+\int_0^{\theta}\frac{d\theta'}{a\Xi\sqrt{\lambda-\sin^2\theta'-\frac{a^2\lambda\cos^2\theta'}{\epsilon L^2}}}+g'(\lambda)\nonumber\\
&=\int_{\chi(r,\lambda)}^{\theta}\frac{d\chi'}{a\Xi\sqrt{\lambda-\sin^2\chi'-\frac{a^2\lambda\cos^2\chi'}{\epsilon L^2}}}+\int_0^{\theta^*}\frac{d\chi'}{a\Xi\sqrt{\lambda-\sin^2\chi'-\frac{a^2\lambda\cos^2\chi'}{\epsilon L^2}}}+g'(\lambda)
\label{Fg}
\end{align}
\end{widetext}
where 
\begin{equation}\label{sq*}
\sin^2\theta^*=\frac{\Xi\lambda}{1-\frac{a^2\lambda}{\epsilon L^2}}
\end{equation}
and the second term in \eqref{Fg}  can be absorbed into $g'(\lambda)$ since it
is independent of $r$ and $\theta$.
\begin{figure}[t]
\centering
\begin{subfigure}{0.5\textwidth}
\includegraphics[width=0.90\linewidth]{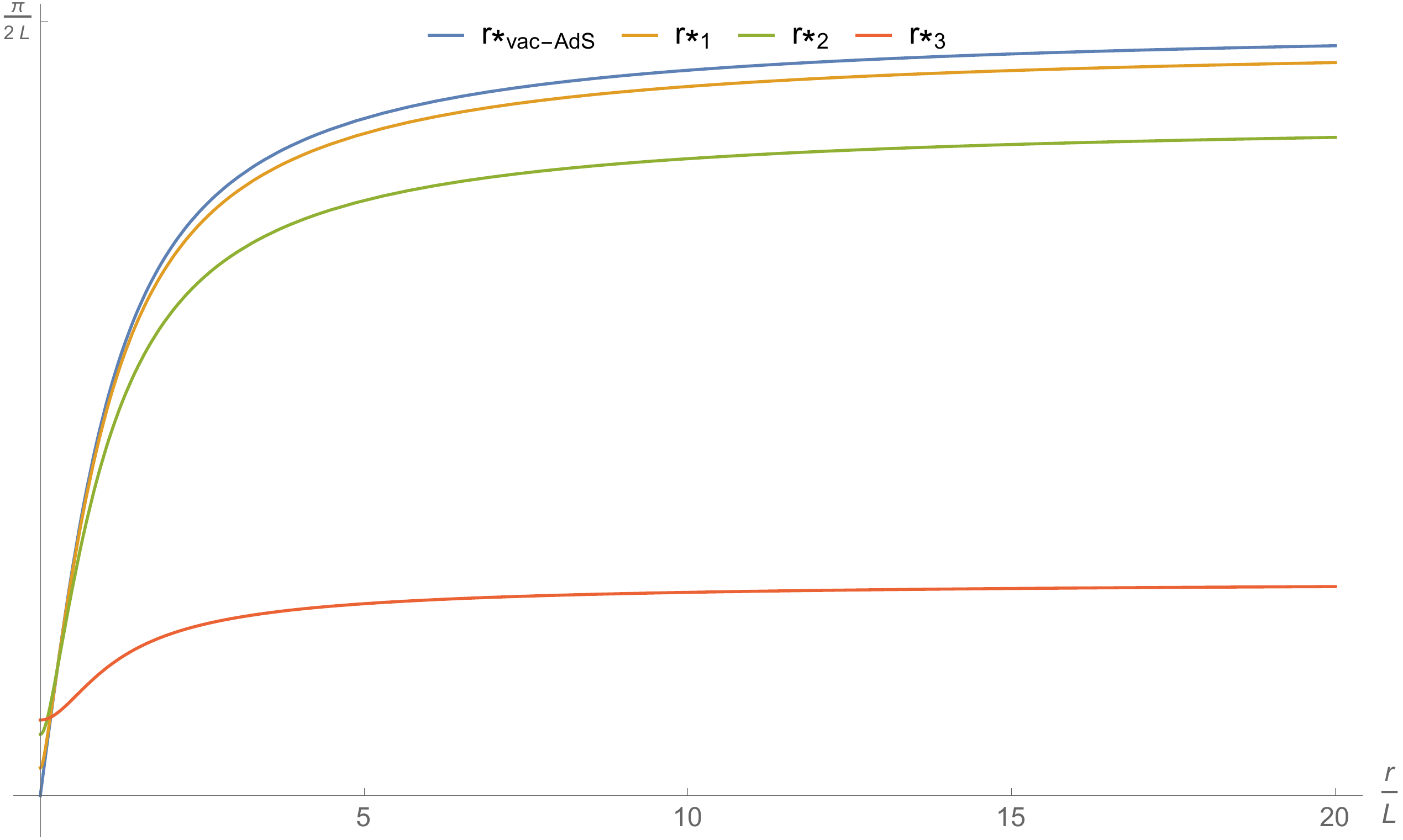}
 \caption{AdS vacuum}
\label{fig1a}
\end{subfigure}
\begin{subfigure}{0.5\textwidth}
 \includegraphics[width=0.90\linewidth]{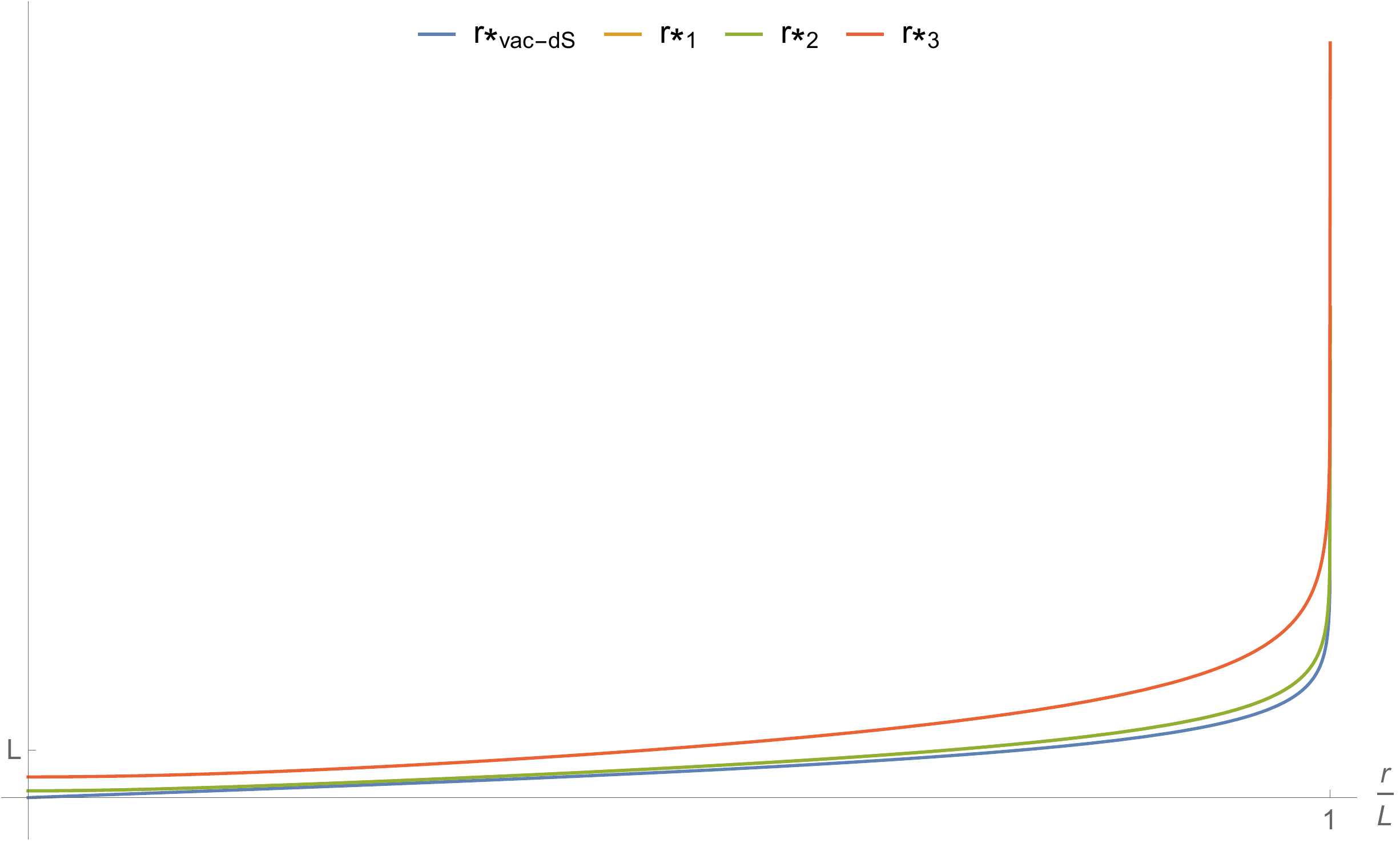}
\caption{dS vacuum}
\label{fig1b}
\end{subfigure}
\captionsetup{justification=centering}
\caption{The tortoise coordinate of (a) vacuum AdS, $r_*=L\tan^{-1}\left(\frac{r}{L}\right)$, and (b) vacuum dS, $r_*=L\tanh^{-1}=\left(\frac{r}{L}\right)$, compared with the numerical solution of \eqref{r*oblate} with $r_1: a/L=0.15$ (orange), $r_2: a/L=0.35$ (green), $r_3: a/L=0.85$ (red) for  $\theta=\frac{\pi}{8}$. Convergence of \eqref{r*oblate} in the limit $a\rightarrow 0$ is implied, as expected from the  transformation \eqref{coordtrans1}-\eqref{coordtrans2}.}
\label{fig1} 
\end{figure}
Choosing $g(\lambda)$ such that the overall constant term in \eqref{Fg} is 0, the constraint $F=0$ now gives 
\begin{equation}
\chi(r,\lambda)=\theta
\label{chitheta}
\end{equation}
From \eqref{r}, this fixes the function $\lambda(r,\theta)$ to be
\begin{equation}\label{lamrq}
\lambda(r,\theta)=\frac{\left(r^2+a^2\right)\sin^2\theta}{r^2\Delta_\theta+a^2\sin^2\theta}
\end{equation}
Integrating \eqref{r**} gives
\begin{widetext}
\begin{align}
r_*&=\int_0^r\frac{Q_0}{\Delta_\chi}dr'+\int_0^\theta\frac{P}{\Delta_\theta}d\theta' =\int_0^{\chi(r,\lambda)}\frac{Q_0^2}{\Delta_r}\frac{d\chi'}{a\Xi\sqrt{\lambda-\sin^2\chi-\frac{a^2\lambda\cos^2\chi}{\epsilon L^2}}}+\int_0^\theta\frac{P}{\Delta_\theta}d\theta'\nonumber\\
&=\int_0^{\chi(r,\lambda)}\frac{a\Xi\lambda(1-\lambda)}{\left(\lambda-\sin^2\chi'\right)\sqrt{\lambda-\sin^2\chi'-\frac{a^2\lambda\cos^2\chi'}{\epsilon L^2}}}d\chi'+\int_0^{\chi(r,\lambda)}\frac{a\Xi\sqrt{\lambda-\sin^2\chi-\frac{a^2\lambda\cos^2\chi}{\epsilon L^2}}}{1-\frac{a^2\lambda\cos^2\chi'}{\epsilon L^2}}d\chi'
\label{r*oblate}
\end{align}
\end{widetext}
and upon substituting $\lambda(r,\theta)$ from \eqref{lamrq} into the result gives $r_* = r_*(r,\theta)$
for (A)dS spacetime.  The result of this integration in \eqref{r*oblate}
can be written in terms of elliptic integrals which, in the limit 
$L\to \infty$ reduce to the flat space case  \cite{Pretorius1998}
\be 
r_* = \sqrt{r^2+a^2\sin^2\theta}
\label{r*flatoblate}
\ee
as we show in the Appendix, where  $\lambda= \sin\theta_*$ from \eqref{sq*}.  In the limit
$a\rightarrow 0$ the expression for $r_*$ 
reduces to the tortoise coordinate in vacuum (A)dS 
\begin{equation}
r_{\text{vac-AdS}}^*(r)=L\tan^{-1}\left(\frac{r}{L}\right)\quad r_{\text{vac-dS}}^*(r)=L\tanh^{-1}\left(\frac{r}{L}\right)
\end{equation}
which we illustrate numerically in figure~\ref{fig1}.

Thus, the surfaces $t\pm r_*=\text{const}$, with $r_*(r,\theta)$ given by \eqref{r*oblate}, are the null hypersurfaces of vacuum (A)dS.  

\section{Properties of axisymmetric null hypersurfaces}
The condition $F=0$ implies that $dF=0$ yielding
\begin{equation}
\mu d\lambda=-\frac{dr}{Q}+\frac{d\theta}{P},\qquad \mu\equiv -\partial_\lambda F
\label{lambda}
\end{equation}
from   \eqref{F}.  In conjunction with  \eqref{r**}, this shows that lines of constant $r_*$ and $\lambda$ are orthogonal, $\nabla r_*.\nabla\lambda=0$, with respect to the intrinsic 2-metric
\begin{equation}\label{2met}
d\sigma^2=\Sigma^2\left[\frac{dr^2}{\Delta_r}+\frac{d\theta^2}{\Delta_\theta}\right]
\end{equation} 
 of the $(t,\phi)$ sections of the Kerr-(A)dS spacetime. Since $\lambda$ is independent of $t$ and $\phi$,  this in turn implies
that  $\lambda$ is constant along the null generators in \eqref{nullgen}
\be
\ell^\alpha \partial_\alpha \lambda = 0    \qquad  n^\alpha \partial_\alpha \lambda = 0
\ee
where $\ell_\alpha = -\partial_\alpha v$ and $n_\alpha = -\partial_\alpha u$.

The 2-metric \eqref{2met} can be written in $(r_*,\lambda)$ coordinates instead of $(r,\theta)$ coordinates. It can easily be checked that
\begin{equation}
d\sigma^2=\frac{1}{\Xi^4R^2}\left[\Delta_r\Delta_\theta {dr_*}^2+P^2Q^2 \mu^2 d{\lambda}^2\right]
\end{equation}
where 
\begin{align}
R^2&\equiv \frac{g_{\phi\phi}}{\sin^2\theta}\nonumber\\
&=\frac{\Delta_\theta{(r^2+a^2)}^2-\Delta_r a^2\sin^2\theta}{\Sigma^2\Xi^2}
\end{align}

We can rearrange \eqref{gtt} so that
\begin{align}
g_{tt}-\frac{g_{t\phi}^2}{g_{\phi\phi}}&=\frac{1}{g^{tt}}\nonumber\\
g_{tt}-\omega_B^2g_{\phi\phi}&=\frac{1}{g^{tt}}\nonumber\\
\frac{\Delta_r-a^2\Delta_\theta \sin^2\theta}{\Sigma^2}+\omega_B^2R^2\sin^2\theta&=\frac{\Delta_r\Delta_\theta}{\Xi^4R^2}
\end{align}
where
\begin{align}
\omega_B&\equiv-\frac{g_{t\phi}}{g_{\phi\phi}}=a\frac{\Delta_\theta(r^2+a^2)-\Delta_r}{\Xi\Sigma^2 R^2}
\label{omegaBB}
\end{align}
is the angular velocity \cite{Thorne1988}. 
This allows us to rewrite the Kerr-(A)dS metric \eqref{Kerr} as
\begin{widetext}
\begin{equation}
ds^2=\frac{\Delta_r\Delta_\theta}{\Xi^4R^2}\left({dr_*}^2-dt^2\right)+R^2\sin^2\theta{\left(d\phi-\omega_Bdt\right)}^2+\frac{P^2Q^2}{\Xi^4R^2} \mu^2 d{\lambda}^2
\label{KerrNew}
\end{equation}
\end{widetext}
 The new coordinates $(t,r_*,\lambda,\phi)$ are better suited to study the null hypersurfaces of Kerr spacetime since the 3-spaces defined by $t\pm r_*=\text{constant}$ are clearly null hypersurfaces,
 and $g^{r_* r_*} = -g^{tt}$.   The degenerate intrinsic metric  is
 \begin{equation}
 dh^2=\frac{P^2Q^2}{\Xi^4R^2} \mu^2d{\lambda}+R^2\sin^2\theta{\left(d\phi-\omega_Bdt\right)}^2
 \label{nullsubmanifold}
 \end{equation}
 on these null hypersurfaces. 
Inversion of the differentials \eqref{r**} and \eqref{lambda} gives
\begin{align}
\Xi^4\Sigma^2R^2dr &=\Delta_rQ\left[\Delta_\theta dr_*-P^2 \mu d{\lambda}\right] \nonumber \\
\Xi^4\Sigma^2R^2d\theta &=\Delta_\theta P\left[\Delta_r dr_*+Q^2 \mu d{\lambda}\right]
\label{drdtheta}
\end{align}

We pause to comment on the construction of quasi-spherical light cones.   These surfaces are those that reduce to the light cones of (A)dS spacetime as $r\to \infty$.  We first note that
insertion of  \eqref{lamrq} into \eqref{sq*} yields
\be
\sin^2\theta^* =  \frac{(r^2+a^2)\sin^2\theta}{r^2+a^2\sin^2\theta}
\ee
which is the same relation as in the asymptotically flat case  \cite{Pretorius1998}, and we have
$\theta^*(r\to \infty,\theta) = \theta$.  Requiring this asymptotic condition 
and the relation  \eqref{sq*} to hold in the Kerr-(A)dS spacetime fixes the function $g'(\lambda)$ in \eqref{F} to yield the relation
\begin{equation}
F(r,\theta,\lambda)= \int_r^\infty \frac{1}{Q(r',\lambda)} dr'  
- \int_\theta^{\theta^*}\frac{1}{P(\theta',\lambda)}d\theta'
\label{Flt}
\end{equation}
The equations of the quasi-spherical hypersurfaces are respectively given by $v = t + r_*(r,\theta) = v_0$
and $u = t - r_*(r,\theta) = u_0$ where $(u_0,v_0)$ are both constants, and $r_* = \rho(r,\theta,\lambda(r,\theta))$, with the function $\lambda(r,\theta)$ determined by setting $F=0$.
 
 \begin{figure}[t]
\centering
\begin{subfigure}{0.5\textwidth}
\includegraphics[width=0.95\linewidth]{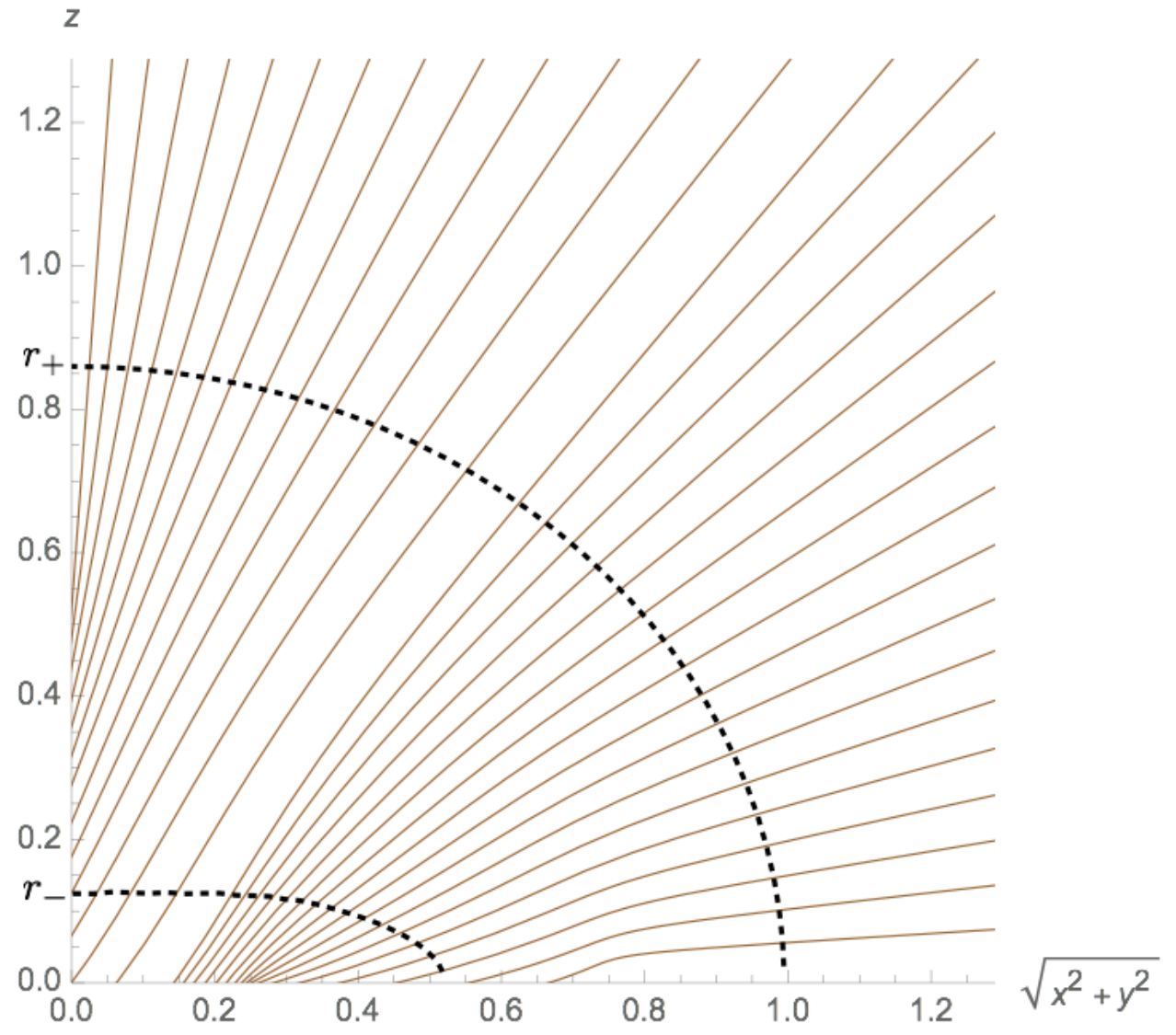}
 \caption{Kerr-AdS}
\label{fig2a}
\end{subfigure}
\begin{subfigure}{0.5\textwidth}
 \includegraphics[width=0.94\linewidth]{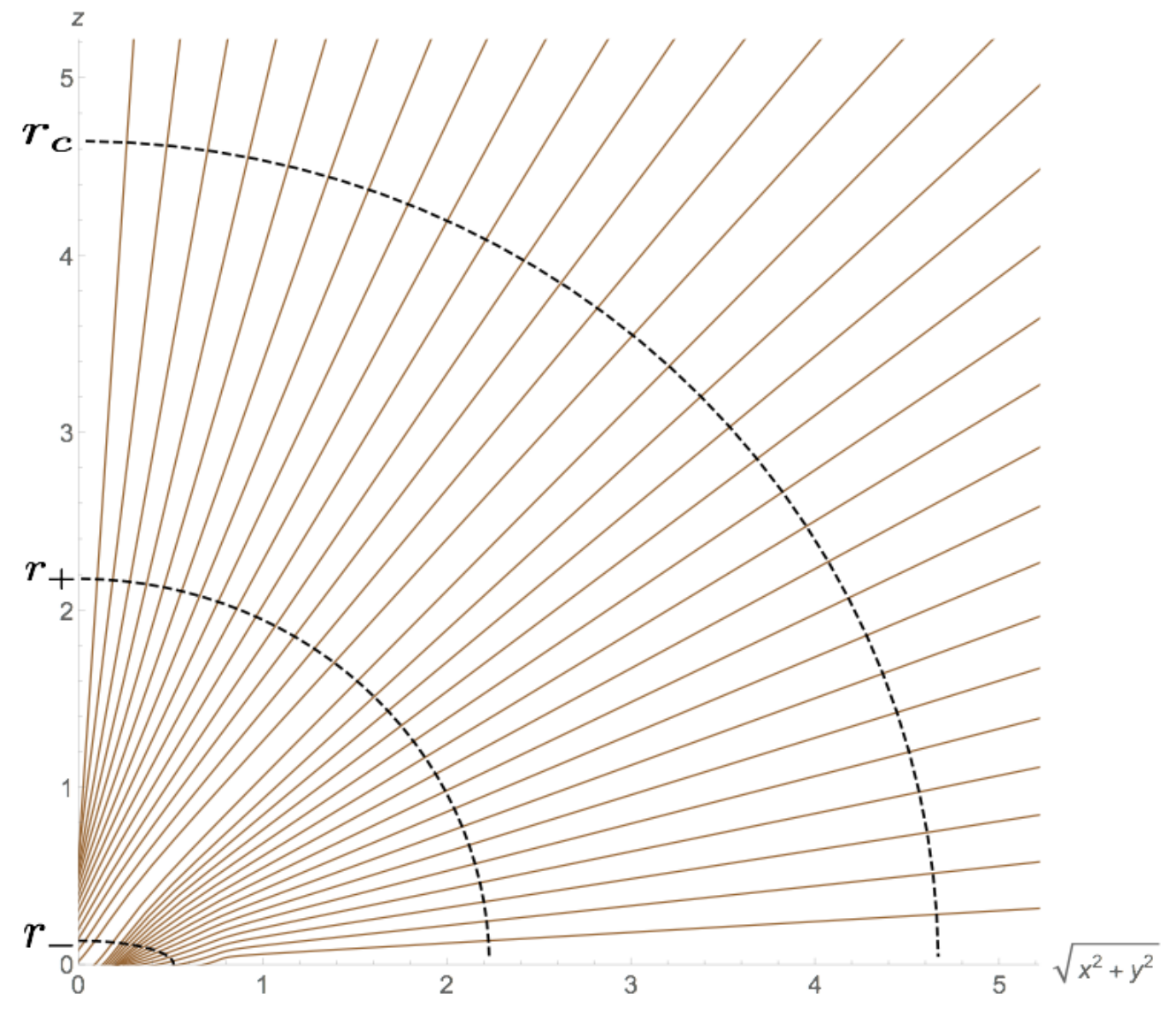}
\caption{Kerr-dS}
\label{fig2b}
\end{subfigure}
\captionsetup{justification=centering}
\caption{Projection of the path of null generators onto the Cartesian plane $(\sqrt{x^2+y^2}=\sqrt{r^2+a^2}\sin\theta,z=r\cos\theta)$ for (a) Kerr-AdS spacetime with $m=L=2a=1$, and (b) Kerr-dS spacetime with $m=2a=1$ and  $L\approx 6.12372$. In both cases, the light sheets do not converge to form a caustic near the black hole. Also shown are the various horizons of the spacetime $r_i$, $i=-,+,c$.}
\label{fig2} 
\end{figure}

 \section{The absence of caustics}

From \eqref{nullsubmanifold}, the condition for the null 3-space to develop a caustic is 
\begin{equation}
\Xi^{-2}\mu PQ\sin\theta=0
\label{caustic}
\end{equation}
In the limit $m\rightarrow 0$, the light cones in vacuum (A)dS spacetime have no caustics. Light cones in the Kerr metric were shown to be free of caustics in \cite{Pretorius1998}. Here, we show that the proof also extends to Kerr-(A)dS spacetimes. 

The proof proceeds by showing that each factor in \eqref{caustic} for $m>0$ increases along an ingoing null generator of fixed $\lambda$ and decreasing $r$, which proves that \eqref{caustic} will not be satisfied in Kerr-(A)dS since it is not satisfied in its $m\rightarrow 0$ limit. 
Initially, when $r\rightarrow\infty$, it is clear from \eqref{r} that $\lambda>0$. From \eqref{lambda} and \eqref{PQ}, $P>0$ increases as $r$ decreases along the generator of fixed $\lambda$. Also, from \eqref{PQ}, it is clear that $Q>Q_0>0$. 

The remaining task is to show that $\mu$ and $\theta$ both increase with increasing $m$, while $r$ and $\lambda$ are kept fixed, that is
\begin{equation}
{\left(\frac{\partial\theta}{\partial m}\right)}_{r,\lambda} > 0 \qquad {\left(\frac{\partial\mu}{\partial m}\right)}_{r,\lambda} > 0
\end{equation}
provided $r>0$ and $\theta_* < \pi/2$. From \eqref{F} and the condition $F=0$, it is straightforward to show that
\begin{equation}
{\left(\frac{\partial\theta}{\partial m}\right)}_{r,\lambda}=a^2\Xi^2\lambda P \int_r^\infty\frac{r'dr'}{Q^3(r',\lambda,m)}
\label{theta}
\end{equation}
Similarly, from \eqref{lambda} and \eqref{F}, 
\begin{align} 
{\left(\frac{\partial\mu}{\partial m}\right)}_{r,\lambda}&=\frac{\partial\mu}{\partial m}+\frac{\partial\mu}{\partial\theta}{\left(\frac{\partial\theta}{\partial m}\right)}_{r,\lambda}\nonumber\\
&=a^2\Xi^2\left(1+\frac{a^2\Xi^2\lambda\Delta_\theta}{2P^2}\right)\int_r^\infty\frac{r'dr'}{Q^3(r',\lambda,m)} \nonumber\\
&\qquad\qquad +\frac{3}{2}a^4\Xi^4\lambda\int_r^\infty\frac{r'\Delta_r(r',m)dr'}{Q^5(r',\lambda,m)}
 \nonumber\\
&= a^4\Xi^4\int_r^\infty\frac{r'}{Q^3(r',\lambda,m)}\frac{\sin^2\theta}{2P^2}dr'  \nonumber\\
&\qquad +\frac{3}{2}a^2\Xi^4\int_r^\infty\frac{r'{(r'^2+a^2)}^2}{Q^5(r',\lambda,m)}dr'
\label{mu}
\end{align}
where the last line follows from \eqref{PQ}.
Both expressions \eqref{theta} and \eqref{mu} are clearly positive regardless of the sign of 
the cosmological constant. Therefore, the condition \eqref{caustic} is never satisfied in Kerr-(A)dS and the null hypersurfaces \eqref{nullsubmanifold} do not develop caustics when propagated toward $r\rightarrow 0$.

In figure~\ref{fig2}, the $\lambda=\text{const.}$ curves were obtained for Kerr-AdS (figure~\ref{fig2a}) and Kerr-dS (figure~\ref{fig2b}) spacetimes by numerically solving the evolution equations of the null generators along ingoing null hypersurfaces
\be
l^\alpha=\frac{dx^\alpha}{d\tau}=-g^{\alpha\beta}\partial_\beta v
\label{levol}
\ee
where $\tau$ is the affine parameter along the null path. In particular, \eqref{levol} yield
\be
\dot{r}=-\frac{Q}{\Sigma^2},\qquad \dot{\theta}=-\frac{P}{\Sigma^2}
\ee
which we numerically integrate to plot the null paths in figure~\ref{fig2}. Also shown are the spacetime horizons: $r_-$, $r_+$, and $r_c$ for Kerr-dS. Figure ~\ref{fig2} shows that no caustics are formed in both spacetimes near the black hole singularity.

As in the asymptotically flat case, we find that if we extend the spacetime to $r<0$, then
null rays moving inward toward the origin in the $r>0$ sheet are deflected outward by the ring singularity
and defocused.  Upon passing through the $r=0$ disc they are refocused and form a caustic in
the $r<0$ region.

\section{Kruskal Coordinates}
Using the transformations
\be
-\frac{dU}{\kappa U} = d(t-r_*)  = du  \qquad  \frac{dV}{\kappa V} = d(t+r_*) = dv
\ee
we can rewrite the metric \eqref{KerrNew} as
\begin{widetext}
\begin{equation}
ds^2=\frac{\Delta_r\Delta_\theta}{\Xi^4R^2} \frac{dU dV}{\kappa^2 UV} +R^2\sin^2\theta{\left(d\phi-\frac{\omega_B}{2\kappa} \left(\frac{dV}{V} -\frac{dU}{U} \right) \right)}^2+\frac{P^2Q^2}{\Xi^4R^2} \mu^2 d{\lambda}^2
\label{KrusNew}
\end{equation}
\end{widetext}
where $\kappa=\kappa_i$ is the surface gravity  
\be
\kappa_i= \frac{d\Delta_r}{dr}\left|_{r=r_i}\right. \qquad  i = +, -, c 
\ee
for the respective outer, inner,  and (for $\epsilon = -1$) cosmological horizon as relevant. Since
\be
UV = - e^{2\kappa r_*}  \qquad  \frac{V}{U} =  - e^{2\kappa t} 
\ee
it is clear that the first term in the metric is regular at the horizon surfaces $U=0$ and $V=0$.
However the second term in \eqref{KrusNew} is not regular on these surfaces.  This defect can
be overcome by defining
\be
\varphi_+ = \phi  +\alpha(r,\lambda)  \qquad \varphi_-  = \phi  - \alpha(r,\lambda)  
\ee
where
\begin{align}
\alpha(r,\lambda)&=a\Xi^3\bigg[\int_0^r\frac{r'^2+a^2}{\Delta_r(r')Q(r',\lambda)}dr'\nonumber\\
&\qquad\qquad -\int_0^\theta\frac{1}{\Delta_\theta(\theta')P(\theta',\lambda)}d\theta'\bigg]
\end{align}
Then, from \eqref{drdtheta}, \eqref{omegaBB}, and \eqref{PQ}, 
\begin{align}
d\alpha&=\omega_Bdr_*-Nd\lambda
\end{align}
where
\begin{widetext}
\begin{align}
N&\equiv a\mu\frac{(r^2+a^2)P^2+Q^2}{\Xi\Sigma^2R^2} -\frac{a^3\Xi^5}{2}\bigg[\int_0^r\frac{r'^2+a^2}{Q^3(r',\lambda)}dr'+\int_0^\theta\frac{1}{P^3(\theta',\lambda)}d\theta'\bigg]
\end{align}
Hence, the metric \eqref{KerrNew} becomes
\begin{equation}
ds^2=\frac{\Delta_r\Delta_\theta}{\Xi^4R^2} \frac{dU dV}{\kappa^2 UV} 
+R^2\sin^2\theta{\left( d\varphi_+-\omega_Bdv+Nd\lambda \right)}^2
+\frac{P^2Q^2}{\Xi^4R^2} \mu^2 d{\lambda}^2
\label{KrusNew1}
\end{equation}
or alternatively 
\begin{equation}
ds^2=\frac{\Delta_r\Delta_\theta}{\Xi^4R^2} \frac{dU dV}{\kappa^2 UV} 
+R^2\sin^2\theta{\left(d\varphi_--\omega_Bdu-Nd\lambda \right)}^2
+\frac{P^2Q^2}{\Xi^4R^2} \mu^2 d{\lambda}^2
\label{KrusNew2}
\end{equation}
\end{widetext}
depending on which sheet is of interest. For example, \eqref{KrusNew1} is regular for the future horizons for both outer and inner horizons of the black hole (or the future cosmological horizon in Kerr-dS), where $\varphi_+$ is constant along each ingoing generator and $v$ is constant along ingoing light sheets.  Alternatively, \eqref{KrusNew2} is regular for the past horizons for both outer and inner horizons of the black hole (or the past cosmological horizon in Kerr-dS), where $\varphi_-$ is constant along each outgoing generator and $u$ is constant along outgoing light sheets.  In both cases the function $N$ is regular.
 
\section{Conclusion} 

We have described a three-dimensional foliation of Kerr-(A)dS spacetimes
in terms of quasi-spherical light cones, extending the construction in the asymptotically flat case
\cite{Pretorius1998}.  We find that both spacetimes are free of caustics for $r>0$. The limit of zero mass of this foliation was taken and we proved that it reduces to light cones in vacuum (A)dS spacetime. As an application, we derived a generalization of Kruskal coordinates for Kerr-(A)dS based on these quasi-spherical light cones.  

These results should prove useful in studies of the AdS/CFT `complexity equals action' conjecture \cite{Brown2016a,Brown2016} for rotating black holes insofar as it provides a useful tool for the construction of null boundaries in these kinds of spacetimes.   It is likewise natural to consider a similar type of analysis to study the properties of light cones in more exotic spacetimes.  Work on these issues is in progress.

 \begin{widetext}
 \section{Acknowledgements}
 This work was supported in part by the Natural Sciences Engineering Research Council. We thank Frans Pretorius for helpful correspondence.
\begin{appendix}
\section{$r_*$ expansion in large $L$ limit}
We derive a series expansion of \eqref{r*oblate} in the large $L$ limit that explicitly reduces to \eqref{r*flatoblate} when $L\rightarrow 0$. First, the integral \eqref{r*oblate} can be written as
\begin{align}
r_*&\equiv\int_0^{\chi(r,\lambda)}I(\chi,\lambda)\nonumber\\
=&\int_0^{\chi(r,\lambda)}\bigg(\frac{a\Xi\lambda(1-\lambda)}{\left(\lambda-\sin^2\chi'\right)\sqrt{\lambda-\sin^2\chi'-\frac{a^2\lambda\cos^2\chi'}{\epsilon L^2}}}+\frac{a\Xi\sqrt{\lambda-\sin^2\chi-\frac{a^2\lambda\cos^2\chi}{\epsilon L^2}}}{1-\frac{a^2\lambda\cos^2\chi'}{\epsilon L^2}}\bigg)d\chi'
\end{align}
Expanding the integrand in powers of $\frac{1}{L}$ gives the series expansion

\begin{align}
I(\chi,\lambda)&=\frac{a\left(\cos^4\chi+2\lambda\cos^2\chi-2\cos^2\chi-\lambda+1\right)}{{\left(\cos^2\chi+\lambda-1\right)}^{3/2}}+\frac{a^3}{2{\left(\cos^2\chi+\lambda-1\right)}^{5/2}}\bigg(2\lambda\cos^8\chi+6\lambda^2\cos^6\chi-7\lambda\cos^6\chi\nonumber\\&+6\lambda^3\cos^4\chi-2\cos^6\chi-14\lambda^2\cos^4\chi+2\lambda^4\cos^2\chi+2\lambda\cos^4\chi-8\lambda^3\cos^2\chi+6\cos^4\chi+2\lambda^2\cos^2\chi\nonumber\\&+7\lambda\cos^2\chi-6\cos^2\chi+2\lambda^2-4\lambda+2\bigg)\frac{1}{\epsilon L^2}+\mathcal{O}\left(\frac{1}{L^4}\right)
\end{align}
Integrating this expression and substituting for \eqref{chitheta} gives
\begin{align}
\int_0^{\chi(r,\lambda)}I(\chi,\lambda)&= \frac{a\sin\theta\cos\theta}{\sqrt{\cos^2\theta+\lambda-1}}-\frac{a^3}{3\sin\theta{\left(\cos^2\theta+\lambda-1\right)}^{3/2}}\bigg[i{(1-\lambda)}^{3/2}\sin\theta{(\cos^2\theta+\lambda-1)}^{3/2}\nonumber\\&\bigg((\lambda+1)E\left(\cos\theta,\sqrt{\frac{1}{1-\lambda}}\right)-\lambda F\left(\cos\theta,\sqrt{\frac{1}{1-\lambda}}\right)\bigg)+\cos\theta(\cos^2\theta-1)\bigg(\lambda\cos^4\theta+(2\lambda^2-\frac{5}{2}\lambda-2)\cos^2\theta\nonumber\\&+\lambda^3-3\lambda^2+2\bigg)\bigg]\frac{1}{\epsilon L^2}+\mathcal{O}\left(\frac{1}{L^4}\right)
\end{align}
where $F(\phi,m)$ is the elliptic integral of the first kind and $E(\phi,m)$ is the elliptic integral of the second kind. Substituting for \eqref{lamrq} and expanding in $\frac{1}{L}$ and simplifying the expression gives
\begin{align}
r_*&=\sqrt{r^2+a^2\sin^2\theta}-\bigg[\frac{r^2(r^2+a^2)}{2\sqrt{r^2+a^2\sin^2\theta}}-\frac{1}{6{\left(r^2+a^2\sin^2\theta\right)}^{5/2}}\bigg(2ir\cos\theta\big((r^2\cos^2\theta-2a^2\sin^2\theta-2r^2)\nonumber\\&E\left(\cos\theta,\frac{\sqrt{r^2+a^2\sin^2\theta}}{r\cos\theta}\right)+(r^2+a^2)\sin^2\theta F\left(\cos\theta,\frac{\sqrt{r^2+a^2\sin^2\theta}}{r\cos\theta}\right)\big)\nonumber\\&-\frac{(r^2+a^2\sin^2\theta)(a^6\sin^4\theta(2\cos^2\theta-5)+r^2a^4(2\cos^6\theta-5\cos^4\theta+12\cos^2\theta-9)-3a^2r^4+r^6)}{a^3}\bigg]\frac{1}{\epsilon L^2}+\mathcal{O}\left(\frac{1}{L^4}\right)
\end{align}
In the limit $L\rightarrow\infty$, this reduces to the tortoise coordinate in flat space written in oblate coordinates \eqref{r*flatoblate}, which is what we wanted to show. 
 \end{appendix}
\end{widetext}

\bibliographystyle{apsrev4-1}
\bibliography{ref.bib}

\begin{thebibliography}{13}%
\makeatletter
\providecommand \@ifxundefined [1]{%
 \@ifx{#1\undefined}
}%
\providecommand \@ifnum [1]{%
 \ifnum #1\expandafter \@firstoftwo
 \else \expandafter \@secondoftwo
 \fi
}%
\providecommand \@ifx [1]{%
 \ifx #1\expandafter \@firstoftwo
 \else \expandafter \@secondoftwo
 \fi
}%
\providecommand \natexlab [1]{#1}%
\providecommand \enquote  [1]{``#1''}%
\providecommand \bibnamefont  [1]{#1}%
\providecommand \bibfnamefont [1]{#1}%
\providecommand \citenamefont [1]{#1}%
\providecommand \href@noop [0]{\@secondoftwo}%
\providecommand \href [0]{\begingroup \@sanitize@url \@href}%
\providecommand \@href[1]{\@@startlink{#1}\@@href}%
\providecommand \@@href[1]{\endgroup#1\@@endlink}%
\providecommand \@sanitize@url [0]{\catcode `\\12\catcode `\$12\catcode
  `\&12\catcode `\#12\catcode `\^12\catcode `\_12\catcode `\%12\relax}%
\providecommand \@@startlink[1]{}%
\providecommand \@@endlink[0]{}%
\providecommand \url  [0]{\begingroup\@sanitize@url \@url }%
\providecommand \@url [1]{\endgroup\@href {#1}{\urlprefix }}%
\providecommand \urlprefix  [0]{URL }%
\providecommand \Eprint [0]{\href }%
\providecommand \doibase [0]{http://dx.doi.org/}%
\providecommand \selectlanguage [0]{\@gobble}%
\providecommand \bibinfo  [0]{\@secondoftwo}%
\providecommand \bibfield  [0]{\@secondoftwo}%
\providecommand \translation [1]{[#1]}%
\providecommand \BibitemOpen [0]{}%
\providecommand \bibitemStop [0]{}%
\providecommand \bibitemNoStop [0]{.\EOS\space}%
\providecommand \EOS [0]{\spacefactor3000\relax}%
\providecommand \BibitemShut  [1]{\csname bibitem#1\endcsname}%
\let\auto@bib@innerbib\@empty
\bibitem [{\citenamefont {Kerr}(1963)}]{Kerr1963}%
  \BibitemOpen
  \bibfield  {author} {\bibinfo {author} {\bibfnamefont {R.~P.}\ \bibnamefont
  {Kerr}},\ }\href {\doibase 10.1103/PhysRevLett.11.237} {\bibfield  {journal}
  {\bibinfo  {journal} {Physical Review Letters}\ } (\bibinfo {year} {1963}),\
  10.1103/PhysRevLett.11.237}\BibitemShut {NoStop}%
\bibitem [{\citenamefont {Newman}\ \emph {et~al.}(1965)\citenamefont {Newman},
  \citenamefont {Couch}, \citenamefont {Chinnapared}, \citenamefont {Exton},
  \citenamefont {Prakash},\ and\ \citenamefont {Torrence}}]{Newman1965}%
  \BibitemOpen
  \bibfield  {author} {\bibinfo {author} {\bibfnamefont {E.~T.}\ \bibnamefont
  {Newman}}, \bibinfo {author} {\bibfnamefont {E.}~\bibnamefont {Couch}},
  \bibinfo {author} {\bibfnamefont {K.}~\bibnamefont {Chinnapared}}, \bibinfo
  {author} {\bibfnamefont {A.}~\bibnamefont {Exton}}, \bibinfo {author}
  {\bibfnamefont {A.}~\bibnamefont {Prakash}}, \ and\ \bibinfo {author}
  {\bibfnamefont {R.}~\bibnamefont {Torrence}},\ }\href {\doibase
  10.1063/1.1704351} {\bibfield  {journal} {\bibinfo  {journal} {Journal of
  Mathematical Physics}\ } (\bibinfo {year} {1965}),\
  10.1063/1.1704351}\BibitemShut {NoStop}%
\bibitem [{\citenamefont {Gunasekaran}\ \emph {et~al.}(2012)\citenamefont
  {Gunasekaran}, \citenamefont {Kubiz?{\'{a}}kb},\ and\ \citenamefont
  {Manna}}]{Gunasekaran2012}%
  \BibitemOpen
  \bibfield  {author} {\bibinfo {author} {\bibfnamefont {S.}~\bibnamefont
  {Gunasekaran}}, \bibinfo {author} {\bibfnamefont {D.}~\bibnamefont
  {Kubiz?{\'{a}}kb}}, \ and\ \bibinfo {author} {\bibfnamefont {R.~B.}\
  \bibnamefont {Manna}},\ }\href {\doibase 10.1007/JHEP11(2012)110} {\bibfield
  {journal} {\bibinfo  {journal} {Journal of High Energy Physics}\ } (\bibinfo
  {year} {2012}),\ 10.1007/JHEP11(2012)110}\BibitemShut {NoStop}%
\bibitem [{\citenamefont {Altamirano}\ \emph {et~al.}(2014)\citenamefont
  {Altamirano}, \citenamefont {Kubizn{\'{a}}k}, \citenamefont {Mann},\ and\
  \citenamefont {Sherkatghanad}}]{Altamirano2014}%
  \BibitemOpen
  \bibfield  {author} {\bibinfo {author} {\bibfnamefont {N.}~\bibnamefont
  {Altamirano}}, \bibinfo {author} {\bibfnamefont {D.}~\bibnamefont
  {Kubizn{\'{a}}k}}, \bibinfo {author} {\bibfnamefont {R.~B.}\ \bibnamefont
  {Mann}}, \ and\ \bibinfo {author} {\bibfnamefont {Z.}~\bibnamefont
  {Sherkatghanad}},\ }\href {\doibase 10.3390/galaxies2010089} {\bibfield
  {journal} {\bibinfo  {journal} {Galaxies}\ } (\bibinfo {year} {2014}),\
  10.3390/galaxies2010089}\BibitemShut {NoStop}%
\bibitem [{\citenamefont {Hawking}\ \emph {et~al.}(1999)\citenamefont
  {Hawking}, \citenamefont {Hunter},\ and\ \citenamefont
  {Taylor}}]{Hawking:1998kw}%
  \BibitemOpen
  \bibfield  {author} {\bibinfo {author} {\bibfnamefont {S.~W.}\ \bibnamefont
  {Hawking}}, \bibinfo {author} {\bibfnamefont {C.~J.}\ \bibnamefont {Hunter}},
  \ and\ \bibinfo {author} {\bibfnamefont {M.}~\bibnamefont {Taylor}},\ }\href
  {\doibase 10.1103/PhysRevD.59.064005} {\bibfield  {journal} {\bibinfo
  {journal} {Phys. Rev.}\ }\textbf {\bibinfo {volume} {D59}},\ \bibinfo {pages}
  {064005} (\bibinfo {year} {1999})},\ \Eprint
  {http://arxiv.org/abs/hep-th/9811056} {arXiv:hep-th/9811056 [hep-th]}
  \BibitemShut {NoStop}%
\bibitem [{\citenamefont {Hawking}\ and\ \citenamefont
  {Reall}(2000)}]{Hawking2000}%
  \BibitemOpen
  \bibfield  {author} {\bibinfo {author} {\bibfnamefont {S.~W.}\ \bibnamefont
  {Hawking}}\ and\ \bibinfo {author} {\bibfnamefont {H.~S.}\ \bibnamefont
  {Reall}},\ }\href {\doibase 10.1103/PhysRevD.61.024014} {\bibfield  {journal}
  {\bibinfo  {journal} {Physical Review D - Particles, Fields, Gravitation and
  Cosmology}\ } (\bibinfo {year} {2000}),\
  10.1103/PhysRevD.61.024014}\BibitemShut {NoStop}%
\bibitem [{\citenamefont {Bredberg}\ \emph {et~al.}(2011)\citenamefont
  {Bredberg}, \citenamefont {Keeler}, \citenamefont {Lysov},\ and\
  \citenamefont {Strominger}}]{Bredberg:2011hp}%
  \BibitemOpen
  \bibfield  {author} {\bibinfo {author} {\bibfnamefont {I.}~\bibnamefont
  {Bredberg}}, \bibinfo {author} {\bibfnamefont {C.}~\bibnamefont {Keeler}},
  \bibinfo {author} {\bibfnamefont {V.}~\bibnamefont {Lysov}}, \ and\ \bibinfo
  {author} {\bibfnamefont {A.}~\bibnamefont {Strominger}},\ }\bibfield
  {booktitle} {\emph {\bibinfo {booktitle} {{String theory: Formal developments
  and applications. Proceedings, ESF Summer School in High Energy Physics and
  Astrophysics, Cargese, France, June 21-July 3, 2010}}},\ }\href {\doibase
  10.1016/j.nuclphysbps.2011.04.155} {\bibfield  {journal} {\bibinfo  {journal}
  {Nucl. Phys. Proc. Suppl.}\ }\textbf {\bibinfo {volume} {216}},\ \bibinfo
  {pages} {194} (\bibinfo {year} {2011})},\ \Eprint
  {http://arxiv.org/abs/1103.2355} {arXiv:1103.2355 [hep-th]} \BibitemShut
  {NoStop}%
\bibitem [{\citenamefont {Pretorius}\ and\ \citenamefont
  {Israel}(1998)}]{Pretorius1998}%
  \BibitemOpen
  \bibfield  {author} {\bibinfo {author} {\bibfnamefont {F.}~\bibnamefont
  {Pretorius}}\ and\ \bibinfo {author} {\bibfnamefont {W.}~\bibnamefont
  {Israel}},\ }\href {\doibase 10.1088/0264-9381/15/8/012} {\bibfield
  {journal} {\bibinfo  {journal} {Classical and Quantum Gravity}\ }\textbf
  {\bibinfo {volume} {15}},\ \bibinfo {pages} {2289} (\bibinfo {year}
  {1998})}\BibitemShut {NoStop}%
\bibitem [{\citenamefont {Brown}\ \emph
  {et~al.}(2016{\natexlab{a}})\citenamefont {Brown}, \citenamefont {Roberts},
  \citenamefont {Susskind}, \citenamefont {Swingle},\ and\ \citenamefont
  {Zhao}}]{Brown2016a}%
  \BibitemOpen
  \bibfield  {author} {\bibinfo {author} {\bibfnamefont {A.~R.}\ \bibnamefont
  {Brown}}, \bibinfo {author} {\bibfnamefont {D.~A.}\ \bibnamefont {Roberts}},
  \bibinfo {author} {\bibfnamefont {L.}~\bibnamefont {Susskind}}, \bibinfo
  {author} {\bibfnamefont {B.}~\bibnamefont {Swingle}}, \ and\ \bibinfo
  {author} {\bibfnamefont {Y.}~\bibnamefont {Zhao}},\ }\href {\doibase
  10.1103/PhysRevLett.116.191301} {\bibfield  {journal} {\bibinfo  {journal}
  {Physical Review Letters}\ }\textbf {\bibinfo {volume} {116}} (\bibinfo
  {year} {2016}{\natexlab{a}}),\ 10.1103/PhysRevLett.116.191301}\BibitemShut
  {NoStop}%
\bibitem [{\citenamefont {Brown}\ \emph
  {et~al.}(2016{\natexlab{b}})\citenamefont {Brown}, \citenamefont {Roberts},
  \citenamefont {Susskind}, \citenamefont {Swingle},\ and\ \citenamefont
  {Zhao}}]{Brown2016}%
  \BibitemOpen
  \bibfield  {author} {\bibinfo {author} {\bibfnamefont {A.~R.}\ \bibnamefont
  {Brown}}, \bibinfo {author} {\bibfnamefont {D.~A.}\ \bibnamefont {Roberts}},
  \bibinfo {author} {\bibfnamefont {L.}~\bibnamefont {Susskind}}, \bibinfo
  {author} {\bibfnamefont {B.}~\bibnamefont {Swingle}}, \ and\ \bibinfo
  {author} {\bibfnamefont {Y.}~\bibnamefont {Zhao}},\ }\href {\doibase
  10.1103/PhysRevD.93.086006} {\bibfield  {journal} {\bibinfo  {journal}
  {Physical Review D}\ }\textbf {\bibinfo {volume} {93}} (\bibinfo {year}
  {2016}{\natexlab{b}}),\ 10.1103/PhysRevD.93.086006}\BibitemShut {NoStop}%
\bibitem [{\citenamefont {Agnese}\ and\ \citenamefont
  {La~Camera}(2000)}]{Agnese:1999df}%
  \BibitemOpen
  \bibfield  {author} {\bibinfo {author} {\bibfnamefont {A.~G.}\ \bibnamefont
  {Agnese}}\ and\ \bibinfo {author} {\bibfnamefont {M.}~\bibnamefont
  {La~Camera}},\ }\href {\doibase 10.1103/PhysRevD.61.087502} {\bibfield
  {journal} {\bibinfo  {journal} {Phys. Rev.}\ }\textbf {\bibinfo {volume}
  {D61}},\ \bibinfo {pages} {087502} (\bibinfo {year} {2000})},\ \Eprint
  {http://arxiv.org/abs/gr-qc/9907030} {arXiv:gr-qc/9907030 [gr-qc]}
  \BibitemShut {NoStop}%
\bibitem [{\citenamefont {Gibbons}\ and\ \citenamefont
  {Volkov}(2017)}]{Gibbons2017}%
  \BibitemOpen
  \bibfield  {author} {\bibinfo {author} {\bibfnamefont {G.~W.}\ \bibnamefont
  {Gibbons}}\ and\ \bibinfo {author} {\bibfnamefont {M.~S.}\ \bibnamefont
  {Volkov}},\ }\href {\doibase 10.1103/PhysRevD.96.024053} {\bibfield
  {journal} {\bibinfo  {journal} {Physical Review D}\ } (\bibinfo {year}
  {2017}),\ 10.1103/PhysRevD.96.024053}\BibitemShut {NoStop}%
\bibitem [{\citenamefont {Thorne}\ \emph {et~al.}(1988)\citenamefont {Thorne},
  \citenamefont {Price}, \citenamefont {Macdonald},\ and\ \citenamefont
  {Detweiler}}]{Thorne1988}%
  \BibitemOpen
  \bibfield  {author} {\bibinfo {author} {\bibfnamefont {K.~S.}\ \bibnamefont
  {Thorne}}, \bibinfo {author} {\bibfnamefont {R.~H.}\ \bibnamefont {Price}},
  \bibinfo {author} {\bibfnamefont {D.~A.}\ \bibnamefont {Macdonald}}, \ and\
  \bibinfo {author} {\bibfnamefont {S.}~\bibnamefont {Detweiler}},\ }\href
  {\doibase 10.1063/1.2811504} {\bibfield  {journal} {\bibinfo  {journal}
  {Physics Today}\ } (\bibinfo {year} {1988}),\ 10.1063/1.2811504}\BibitemShut
  {NoStop}%
\end{thebibliography}%

\end{document}